\documentclass[pdflatex,sn-mathphys-num]{sn-jnl}

\usepackage{graphicx}%
\usepackage{multirow}%
\usepackage{amsmath,amssymb,amsfonts}%
\usepackage{amsthm}%
\usepackage[title]{appendix}%
\usepackage{xcolor}%
\usepackage{textcomp}%
\usepackage{manyfoot}%
\usepackage{booktabs}%
\usepackage{algorithm}%
\usepackage{algorithmicx}%
\usepackage{algpseudocode}%
\usepackage{listings}%

\theoremstyle{thmstyleone}%
\theoremstyle{thmstyletwo}%

\theoremstyle{thmstylethree}%

\begin{document}

\title[Article Title]{On the generation of astrophysically-relevant intermittent magnetic turbulence in the laboratory}


\author*[1,2]{\fnm{Itamar} \sur{Cohen}}\email{Itamar.oarb@gmail.com}
\author[2,3]{\fnm{Weipeng} \sur{Yao}}
\author[4]{\fnm{Archie F. A.} \sur{Bott}}
\author[5]{\fnm{Sophia N.} \sur{Chen}}
\author[2]{\fnm{Nikola} \sur{Mirkovic}}
\author[6]{\fnm{Jerome} \sur{Béard}}
\author[7]{\fnm{Petrisor Gabriel} \sur{Bleotu}}
\author[7]{\fnm{Georgiana} \sur{Giubegal}}

\author[8]{\fnm{Anda-Maria} \sur{Talposi}}
\author[2]{\fnm{Yoav} \sur{Heller}}
\author[9,10,11]{\fnm{Clément} \sur{Lacoste}}
\author[10]{\fnm{Patrizio} \sur{Antici}}
\author[12]{\fnm{Damiano} \sur{Caprioli}}
\author[10]{\fnm{Emmanuel} \sur{D'Humieres}}
\author[7,8]{\fnm{Victor} \sur{Malka}}
\author[13]{\fnm{Alexandre} \sur{Marcowith}}
\author[7]{\fnm{Ovidiu} \sur{Tesileanu}}
\author[14]{\fnm{Mateusz} \sur{Ruszkowski}}
\author[15]{\fnm{Philipp} \sur{Kempski}}
\author[16]{\fnm{Olga} \sur{Alexandrova}}
\author[1,2]{\fnm{Julien} \sur{Fuchs}}\email{Julien.fuchs@technion.ac.il}

\affil*[1]{\orgdiv{Department of Physics}, \orgname{Technion}, \orgaddress{\city{Haifa}, \postcode{32000}, \country{Israel}}}
\affil[2]{LULI - CNRS, CEA, UPMC Univ Paris 06 : Sorbonne Universit\'e, Ecole Polytechnique, Institut Polytechnique de Paris - F-91128 Palaiseau cedex, France}
\affil[3]{Sorbonne Université, Observatoire de Paris, Université PSL, CNRS, LUX, 75005 Paris, France}
\affil[4]{Department of Physics, University of Oxford, Oxford OX1 3PU, United Kingdom}
\affil[5]{Light Stream Labs LLC, USA, Palo Alto, CA 94306}
\affil[6]{Laboratoire National des Champs Magnétiques Intenses, LNCMI-CNRS, EMFL, Université Grenoble-Alpes, Université Toulouse 3, INSA Toulouse, F-31400 Toulouse, France}
\affil[7]{ELI-NP, “Horia Hulubei” National Institute for Physics and Nuclear Engineering, 30 Reactorului Street, Măgurele RO-077125, Romania}
\affil[8]{Department of Physics of Complex Systems, Weizmann Institute of Science, Rehovot 7610001, Israel}
\affil[9]{CELIA, University of Bordeaux-CNRS-CEA, Talence F-33405, France}
\affil[10]{INRS EMT, Varennes J3X 1P7, Canada}
\affil[11]{Joint Centre for Extreme Photonics, National Research Council and University of Ottawa, Ottawa, Ontario, Canada}
\affil[12]{Department of Astronomy and Astrophysics, The University of Chicago, IL 60637, USA}
\affil[13]{Laboratoire Univers et Particules de Montpellier CNRS/Université de Montpellier, Place E. Bataillon, 34095 Montpellier, France}
\affil[14]{Department of Astronomy, University of Michigan, 1085 S. University Ave., 323 West Hall, Ann Arbor, MI, 48109-1107, USA}
\affil[15]{Instituto Superior Técnico, Universidade de Lisboa, Av Rovisco Pais, 1049-001 Lisboa, Portugal}
\affil[16]{LIRA, Observatoire de Paris, Université PSL, Sorbonne Université, Université Paris Cite, CY Cergy Paris Université, CNRS, 92190 4 Meudon, France}




\abstract{Intermittent magnetic turbulence, namely the presence of non-ordered and clusterized fields, is a ubiquitous phenomenon in space and astrophysical plasmas. It is currently understood that it plays a crucial role in the dynamics of astrophysical systems at all scales, from influencing the evolution of the cosmos as a whole to governing local particle acceleration. While there is direct evidence of turbulence in the solar wind, and despite  progress obtained through multi-wavelength observations, most of our knowledge of it outside the solar system derives from indirect evidence, through modeling. Here we show that magnetic turbulence, that quantitatively matches that measured in space, can be reproduced in the laboratory. Starting from an homogeneous magnetized plasma, we randomly perturb it using a
speckled laser beam. Using proton radiography, we can follow the development and 
quantitatively characterize 
the
produced  intermittent turbulence from its inception. 
Such a platform represents a significant step forward enabling progress on our understanding of a wide variety of astrophysical phenomena, from particle acceleration  to transport of cosmic-rays through the interstellar medium.}


\maketitle


Electromagnetic (EM) turbulence \cite{Schekochihin2007}, in general, is the natural state of many space and astrophysical plasmas, such as the solar wind \cite{Alexandrova2013}, interstellar medium \cite{Lee2020}, and galaxy clusters \cite{Sur2019}. It is thought to shape the dynamics of astrophysical systems at all scales \cite{Popova2023}, regulating processes such as star formation \cite{
Moon2025} or angular-momentum transport in accretion flows \cite{Balbus1998}. Magnetized turbulence also plays a key role in the acceleration and propagation of high-energy charged particles known as cosmic rays (CRs) \cite{Blasi2013}. 


Quantitative characterization of magnetic turbulence, on scales both larger and smaller than particle gyroradii, is possible through direct sampling  by satellites \cite{Alexandrova2013} or through observations of relatively close-by phenomena such as solar flares \cite{Abramenko2003,Malapaka2013}. However such direct measurements are limited to the plasma present in our solar system. For extra-solar systems, there is strong indirect evidence of magnetic turbulence from a variety of observational channels. \textit{The Big Power Law in the Sky} of electron density fluctuations in the interstellar medium suggests that turbulence is not only present, but also spans an impressive range (over 10 orders of magnitude) of spatial scales \cite{Armstrong1995}. Signatures of magnetized turbulence are also present in the intracluster medium of galaxy clusters, as evidenced by measurements of density fluctuations \cite{Zhuravleva2019} and velocity structure functions \cite{Li2020}. 

Due to the importance of the question of turbulence 
in
plasmas, and the aforementioned difficulties linked with direct observations of space plasmas, this has motivated
significant efforts by many groups to perform laboratory experiments to generate and characterize
space and astrophysical-relevant turbulence in plasmas. 
This has been done e.g., with plasma tanks \cite{Howes2012} or laser produced
plasmas \cite{Zhong2026}. Using lasers, recent efforts allowed characterizing fluid turbulence \cite{Rigon2021} as well as
electromagnetic (EM) turbulence. The latter could be done using small lasers \cite{Chatterjee2017
}, but these
are only able to produce turbulent plasmas over 10s of $\mu m$ spatial domain. To produce much larger
($>mm$) turbulent plasmas, higher-energy (using kJ of energy) lasers  were used \cite{Bott2021
}. In these experiments, seeding of the turbulence was done by using meshes
that broke in bits the inflows and the associated Biermann-battery \cite{Campbell2020} and/or seed fields. All these methods have in common that they induce the growth of magnetic fields (B), which are concomitantly rendered turbulent.

Here we show that large-amplitude (with  $(\delta B/B)_{max}$ $\sim$  0.5), astrophysical-relevant (see Appendix \ref{secA2}), magnetic turbulence  over large volume ($>mm$) can be produced using the alternative method of rendering turbulent an already established large-strength magnetic field. The advantage of doing so is that, by requiring  much-less input laser energy and hence by not requiring the use of large-scale facilities, it allows to obtain many snapshots into the produced turbulence, leading to (1) the temporal dynamics of the onset of the turbulence, which could only up to now be simulated, but not measured. A further advantage, related to the large volume of turbulence that can be sampled, is that (2) we can characterize the intermittent nature of the induced turbulence to a much higher degree than achievable up to now (i.e. with  
structure functions  being characterized up to  7-8 moments \cite{Abramenko2003}). Such compact and tunable magnetic turbulence not only allows for repeatability and parameter scanning of the characteristics of magnetic turbulence in plasmas, but it opens the door to probing the fine-scale physics that is inaccessible to current astronomical observations in the domain of turbulence interaction with a wide variety of phenomena, e.g. shocks or particle acceleration and transport. 

We start by imposing a strong, homogeneous
magnetization (see Methods) onto a collisionless plasma (see Appendix \ref{density_temp}), and break it down to a stochastic structure. For this, a high-power multi-speckled laser \cite{
Garnier1999-tu} (see Methods) is directed into the homogeneously magnetized plasma, as illustrated in Fig.~\ref{ExperimentalSetup}. The speckled structure of the laser induces localized plasma heating (see Appendix \ref{density_temp}), which in turn induces a randomization of the magnetic field lines, as illustrated in Fig.~\ref{ExperimentalSetup}b. Note that, due to the non-linear coupling between the laser and the plasma, the input speckled structure of the laser is itself quickly and dynamically randomized \cite{Fuchs2001,Malka2003}, and thus no deterministic imprint of the laser is imparted onto the plasma. The advantage of our setup is that it allows direct control of the generated $\delta B/B$, i.e. the ratio of the stochastic component of the magnetic to the guide field, through the local thermal pressure imposed onto the plasma by the laser speckles. Observations suggest that astrophysical turbulence spans a wide range of regimes, from $\delta B/B \ll 1$ to $\delta B/B \gg 1$ \cite{
Moseley2021}, which strongly motivates a controlled way of accessing both regimes in laboratory turbulence. This is particularly important for the study of cosmic ray propagation and acceleration, as recent theoretical work has shown that the physics of their transport and energization is significantly different in $\delta B/B \gtrsim 1$ and $\delta B / B <1$ turbulence \cite{
Lemoine2023, Kempski2023, Kempski2025, Sebastian2025}. 

The stochastic magnetic fields generated in the experiment are characterized using proton radiography \cite{Schaeffer2023} (see Methods). This probe uses laminar and ps-duration (at the source) but broadband MeV-energy protons that are directed into the turbulent plasma and are collected onto a stack of films in the exit, see Fig.~\ref{ExperimentalSetup}a. The probing protons are non-perturbative due to their low density when they cross the plasma region \cite{Mima2018}, as the proton source target stands far away from it (5 cm, see In Appendix~\ref{background_p}).   This diagnostic allows to measure the field perturbations that altered the probing protons trajectories. Due to the time-of-flight dispersion the  probing protons are subject to before reaching the plasma, the protons collected on each film in the stack, and which are associated to a narrow energy interval, probe the plasma at a different time. Further, we can vary on different shots the time at which the protons are sent in the plasma, with respect to its irradiation by the speckled laser beam. Overall, it allows us to retrieve time-resolved maps of the line-integrated magnetic field structure in the plasma, as a function of time (see Methods). In Appendix~\ref{secA - BvsE} we show that the recorded deflections match those generated by magnetic fields rather than by electric fields. As detailed below, the characterization of the magnetic fluctuations is found to match well that measured in the solar wind by direct satellite sampling, as well that inferred from solar flare data, in terms of spectrum, amplitude and intermittency. The seeded turbulence is found to reach high levels of fluctuations, i.e. with  $(\delta B/B)_{rms}$ around 0.5 and $(\delta B/B)_{rms}$ around 0.2, i.e., with similar level as inferred to be the case around SNRs \cite{Sapienza2022}.

\begin{figure*}
\centering
\includegraphics[trim={0cm 0cm 0cm 0cm},clip, width=0.95\textwidth]{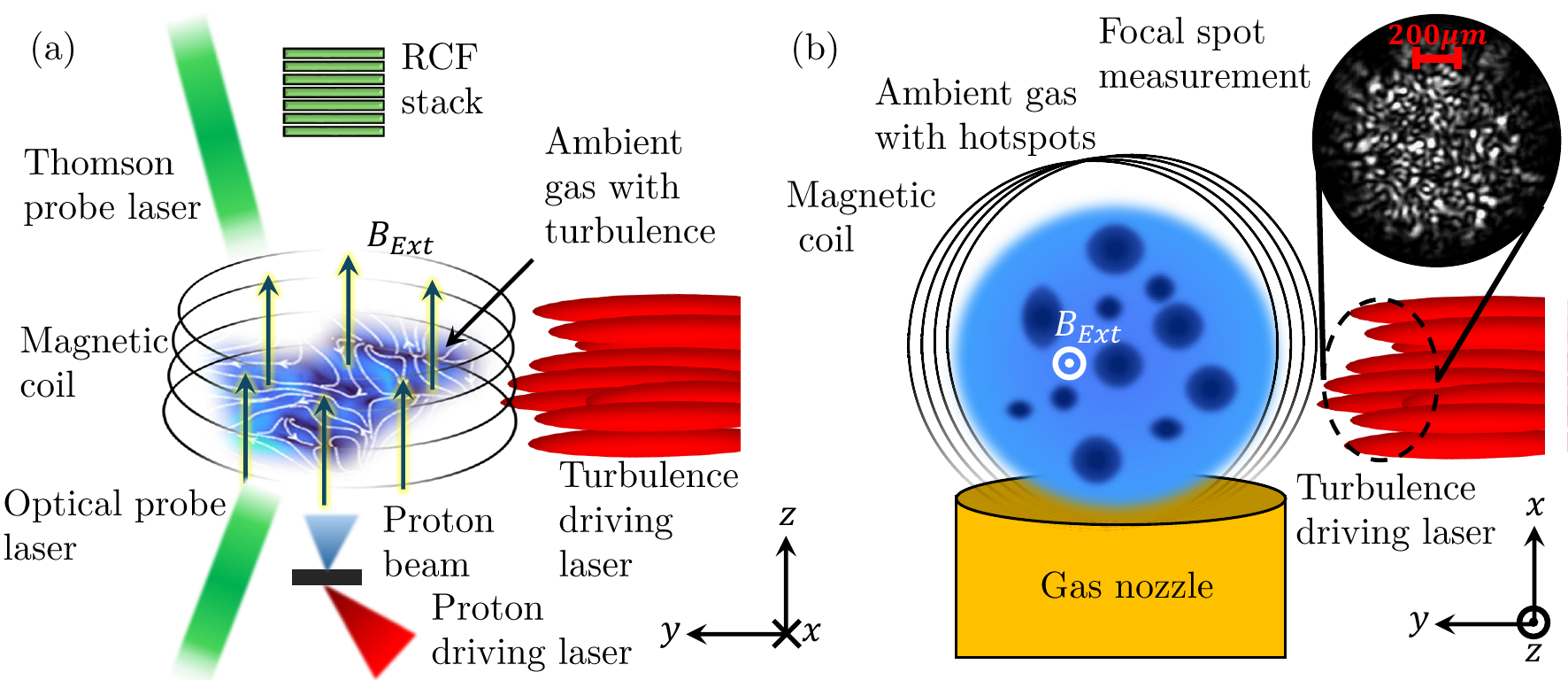}
\caption{\label{ExperimentalSetup} \textbf{Experimental setup.} a) Top and b) side views. An ambient plasma (produced by a pulsed hydrogen gas jet) is positioned inside a coil (having  openings of diameter 1 cm). The coil generates a background magnetic field of $10$~T \cite{Albertazzi2013} along the z-axis. Turbulence is driven by propagating (along the y-axis and through the plasma), a laser having randomized laser speckles (having average diameter of 20~µm). High-energy (MeV) protons, generated using an ultra-short laser pulse, propagate along the external magnetic field to probe the turbulent plasma magnetic field, and are collected onto a stack of radio-chromic films (RCFs). The plasma is also probed by two additional lasers, illustrated in green in panel (a), to perform temperature and density measurements based on Thomson scattering and interferometry (see Methods), respectively. }
\end{figure*} 

\begin{figure*}[hbtp]
\centering
\includegraphics[trim={0.33cm 0cm 2cm 0cm},clip, width=1\textwidth]{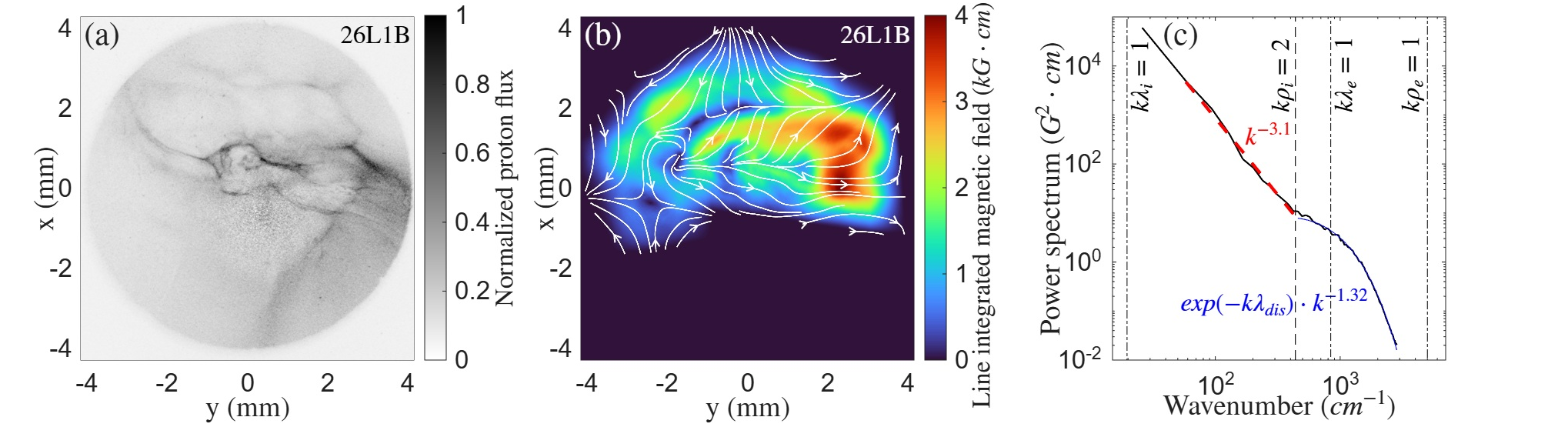}
\caption{\label{TurbuExample} \textbf{Proton deflectometry analysis of controlled magnetic turbulence generation.} (a) proton radiography of magnetic fragmentation generated by having a randomized 5 ns-duration laser beam propagate within a low-density ($5\cdot10^{17}$~$cm^{-3}$) plasma embedded in a strong ($10$ T) external magnetic field (aligned along the z-axis). (b) The path-integrated magnetic field map corresponding to (a), as retrieved using the PROBLEM code \cite{Bott2017}; white arrows represent reconstructed magnetic field lines. (c) The magnetic field power spectra with different power law fittings in different regions. The vertical dashed lines mark relevant plasma scales. Note that, as the probing proton source size is determined by the laser focal spot \cite{Schaeffer2023}, which is $\sim6$ $\mu m$, the cutoff associated with the resolution of the measurement corresponds to a very large wavenumber ($5000$ ${cm}^{-1}$).}
\end{figure*}

Fig.~\ref{TurbuExample}a presents a typical proton deflectometry image (see Methods) that results from the probing proton having propagated through the turbulent magnetic field.  The image corresponds to probing of turbulence at $t=2.77$ ns after the start of the plasma irradiation by the speckled laser beam. Without an external magnetic field initially imposed on the plasma, but with the ambient medium, 
the film collecting the probing protons shows no deflections, i.e. the proton beam profile is very smooth (see Appendix \ref{background_p}). Similarly, with the external magnetic field, but in absence of the speckled laser beam, no deflections are recorded onto the probing proton beam. However, when having both the external  magnetic field and the speckled laser beam applied onto the plasma, it is
clear that the probing proton pattern becomes complex, with a large dark region in the middle surrounded
by several “net”-shaped structures.
An analysis of the proton deflections, in the region which had turbulence seeded in it, using the PROBLEM algorithm (see Methods) was performed to reconstruct the path-integrated (along the direction of the applied field) magnetic field map \cite{Bott2017}.
The path-integrated magnetic field map is shown in Fig.~\ref{TurbuExample}b.

The power spectrum of the induced magnetic turbulence, calculated from the line-integrated magnetic field map shown in Fig.~\ref{TurbuExample}b, and in the same manner as in Ref.\cite{Bott2019},  is shown in  Fig.~\ref{TurbuExample}c. It corresponds to a probing time of $t=2.77$~ns after the start of the speckled laser beam irradiation. The vertical dashed lines mark the different ion (i) and electron (e) plasma scales in the plasma, with $\rho_{e/i}$ being the Larmor radius and $\lambda_{e/i}$ being the plasma inertial length, respectively. We can observe a clear break in the magnetic field power spectrum between the ion and electron scales. The spectrum was thus fitted differently in these two different regions: (1) the ion scale characterized by $k\rho_i<2$, where the spectrum  was fitted by a power law $k^{-p_i}$, with $p_i$ being the power index, and 
(2) the electron scale characterized by $k\rho_i>2$, where the spectrum  was fitted by $exp(-k\lambda_{dis})\cdot k^{-p_e}$, in a similar manner as in solar wind plasmas \cite{Alexandrova2009,Alexandrova2013} (where the index "dis" stands for dissipation). Mind that $\lambda_{dis}$ differs from the inertial lengths in the plasma. We can observe that these two fits are in excellent agreement with the data. In order to decrease the number of free parameters, we fitted all the power spectra
with a single value for the power index $p_e$=1.32 and used as a free parameter to all the spectra the dissipation length $\lambda_{dis}$. For the ion scale part of the spectrum, we remark that, as in solar wind turbulence at the same ion scales, the spectral index $p_i$ of the turbulent spectrum is in the range of 2–4 \cite{
Alexandrova2009}.
Regarding the transition to the electron-scale region, we then observe, also as recorded in  the solar-wind downstream of the Earth’s bow shock, that the turbulence spectrum changes its shape around $k\lambda_e \simeq k\rho_E \sim 1$ \cite{Alexandrova2008}. We note that such a break was observed as well in the upstream solar wind that is magnetically connected to the bow shock \cite{Sahraoui2009}.


\begin{figure}[hbtp]
\centering
\includegraphics[trim={0cm 0cm 0cm 0cm},clip, width=0.95\textwidth]{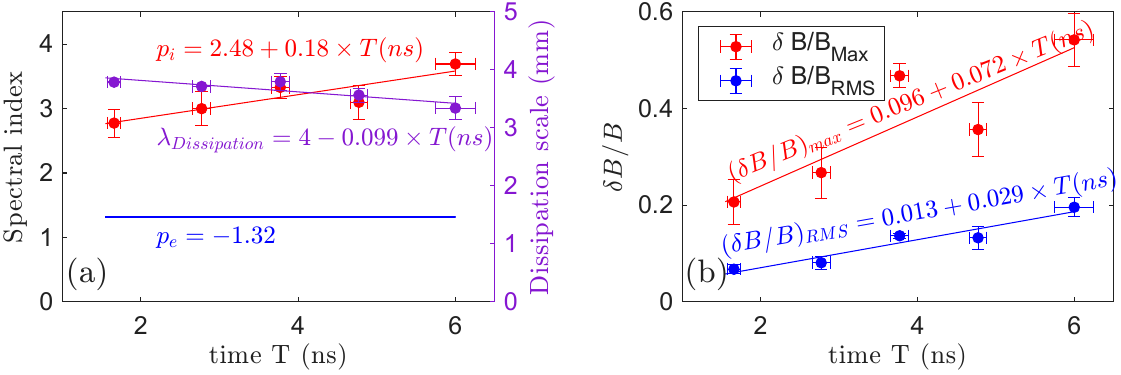} 
\caption{\label{Power Scaling} \textbf{Scalings of the turbulence power spectrum and turbulent magnetic field fluctuations as a function of time.} (a) The spectral indexes (as defined in Fig.~\ref{TurbuExample}) of the magnetic field power spectrum, as fitted in the ion and electron regions, respectively, and as measured over time (time 0 corresponds to the start of the plasma irradiation by the speckled laser beam - the points are measured on different shots, by changing the time of the probing with respect to the arrival of the speckled laser beam). The spectrum in the ion scale region, i.e. corresponding to $k\rho_i<2$, was fitted a power law $k^{-p_i}$. The spectrum in the  electron scale region, i.e. corresponding to $k\rho_i>2$,  was fitted by a weighted power law  $exp(-k\lambda_{dis})\cdot k^{-p_e}$. For this fit, the spectral power $p_e$ was kept constant. The power $p_i$ and $\lambda_{dis}$ exhibit a linear  temporal dependence.
(b) RMS and maximum values of the magnetic field fluctuations, $\delta B/B$, as a function of time. The lines are a linear parameterization.
}

\end{figure}

Similar magnetic field maps and spectra were measured at different times, on different shots, and by changing the time of arrival of the probing protons compared to the turbulence-seeding laser
; they are shown in Appendix \ref{secA1}. They could be fitted similarly as the example discussed above, and which corresponds to a probing time of $t=2.77$~ns. The fitting results of the spectrum obtained at  different probing times are shown in Fig.\ref{Power Scaling}a. Measurements that were close in time were averaged to reduce any effects of errors in the analysis due to possible localized caustics present in the probing proton beam (see Methods). The error bars are calculated from the standard deviation of the averaged data. The results show that the spectrum slopes evolve slowly over 
time, at least more slowly than 
the amplitude of the magnetic field, which is strongly growing over time, as shown below.

To further quantify the turbulence, we measure the growth of the amplitude of the magnetic fluctuations over time. For this, we calculated for each magnetic field map the RMS and the maximum values of the magnetic flied fluctuations, relative to the externally imposed magnetic field. To estimate the amplitude of the magnetic field from the line integrated magnetic field (which is the B field multiplied by the length along the projected view), we divided the results by a typical length $l_{typical}=\sqrt{l_pl_B}$ calculated for each magnetic field map \cite{Bott2017}, which takes into account the length the probing protons propagated through the plasma $l_p\sim3$~mm (inferred from the vertical size of the turbulence  region in Fig.~\ref{TurbuExample}b), and the stochastic length of the path-integrated magnetic field $l_B~\sim$~0.5~mm (calculated from using the power spectrum as detailed in Ref.~\cite{Bott2017
}). The results are shown in Fig.\ref{Power Scaling}b, which shows that the amplitude of the field is strongly growing linearly with time, \textcolor{black}{ even though the  turbulence-seeding laser pulse ended at $t=5$~ns
. We observe that the turbulence can still grow after 5 ns, likely due to energy exchange with the magnetic guide field.} 

\begin{figure}[hbtp]
\centering
\includegraphics[trim={0 0 0 0},clip, width=0.95\textwidth]{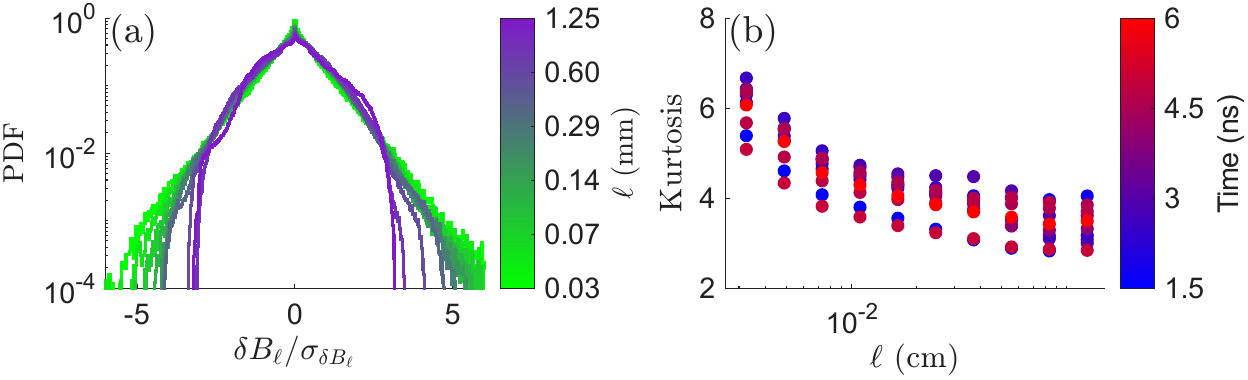} 
\caption{\label{PDF_example} \textbf{Characterization of the intermittency of the turbulent magnetic field.} (a) Probability distribution function (PDF) of the magnetic field variation $\delta B_\ell = (\vec{B}(\vec{r}+\vec{\ell})-\vec{B}(\vec{r}))\cdot \frac{\vec{\ell}}{\ell}$ for the magnetic field map shown in Fig.~\ref{TurbuExample}b. The PDF is calculated for different scales, starting from a slightly larger value than the seeding hot spots within the speckle laser, which are 20~$\mu$m in size. $\sigma_{\delta B_\ell}$ corresponds to the one-sigma deviation in the distribution of $\delta B_\ell$. (b) Kurtosis of the PDF of the magnetic field fluctuations (such as the one shown in (a)), as a function of time and distance $\ell$. The dots correspond to the measured kurtosis.}
\end{figure}

While the power spectra, such as the one shown in Fig.~\ref{TurbuExample}, tell us how the overall variance or energy is distributed across spatial scales, it 
only captures second-order statistics. This type of analysis allows to identify inertial ranges and characteristic scales, but does not easily allow to discern intermittency and the presence of coherent structures.
Probability distribution functions (PDFs), on the other hand, allow to access the full distribution of fluctuation amplitudes at given scales. They allow to demonstrate non-Gaussian behavior and the presence of rare events. For each line-integrated magnetic field map, we calculated the PDF of the correlation in B at various distances, normalized by the standard deviation of the distribution. For this calculation, we randomly picked two sets of $10^9$ points at different positions within the turbulence region.  We then evaluated $\delta B_\ell = (\vec{B}(\vec{r}+\vec{\ell})-\vec{B}(\vec{r}))\cdot \frac{\vec{\ell}}{\ell}$ for all the points, and categorized these values for different values of $\ell$, where $\vec{\ell}$ is the vector connecting the two randomly picked points, $\vec{r}$ is the 2D position vector, and $\vec{B}$ is the 2D projected magnetic field. For these reasons, we choose to weight the increments of $\delta B_\ell$ with a product of $\cdot \vec{\ell}$ to stay closer to the standard structure-function definition used in turbulence \cite{Malapaka2013}, since the magnetic field measurements are path integrated and projected on a plane.
The results of such probability density function (PDF) calculations are shown  in Fig.\ref{PDF_example}a. We can clearly see that, when going to smaller and smaller scales $\ell$, the PDFs have growing wings and deviate more and more from Gaussian distributions that are characteristic of random fluctuations, and thus indicate the presence of intermittent structures within the magnetic turbulence.

To evaluate quantitatively how non-Gaussian the distributions are at various scales, we calculate the kurtosis (i.e. the forth moment divided by the standard deviation squared, $ \langle \delta B_\ell^4\rangle / \langle \delta B_\ell^2\rangle^2$) of the distributions. This is shown, as a function of time and distance $\ell$, in Fig.\ref{PDF_example}b. This allows us to quantify the intermittency in a similar manner to the detection of vortexes in the solar wind \cite{Sorriso1999}. Also, it allows us to quantify how well-developed is the turbulence is as it departs from a Gaussian random noise. The results show that at large scales $\ell\sim1$~mm the kurtosis converges to 3, as  expected since a kurtosis of 3 corresponds to a Gaussian distribution \cite{Roberts2022}. At small scales, i.e. for $\ell\lesssim 0.1$~mm, the kurtosis is larger than 3 and peaks at $\sim$6 at the smallest scale, in a similar manner to that observed in the solar wind \cite{Roberts2022}. The kurtosis is seen to remain stable over time throughout the probing times of the experiment.


Since the PDFs, such as the one shown in Fig.\ref{PDF_example}a,  are hard to compare across scales and require a large amount of data to converge (especially in the tails), we also computed the structure functions of the distributions, in order to 
quantitatively 
describe how turbulence evolves with scale.
Structure functions sit in between the description of power spectra and PDFs, and combine many of their advantages. A structure function is a moment of the PDF of field differences at a given separation, evaluated as a function of that separation. In practice, this provides a compact, scale-by-scale summary of the fluctuation statistics. By examining how structure functions vary with separation, we can directly quantify how turbulent fluctuations grow or shrink with scale and whether the scaling is simple or dominated by intermittent events.
An important advantage of structure functions is that they allow us to go beyond second-order statistics in a controlled way. Since the second-order structure function is directly related to the power spectrum, this approach does not discard spectral information but rather generalizes it.

\begin{figure*}[hbtp]
\centering
\includegraphics[trim={0cm 0cm 0cm 0cm},clip, width=0.95\textwidth]{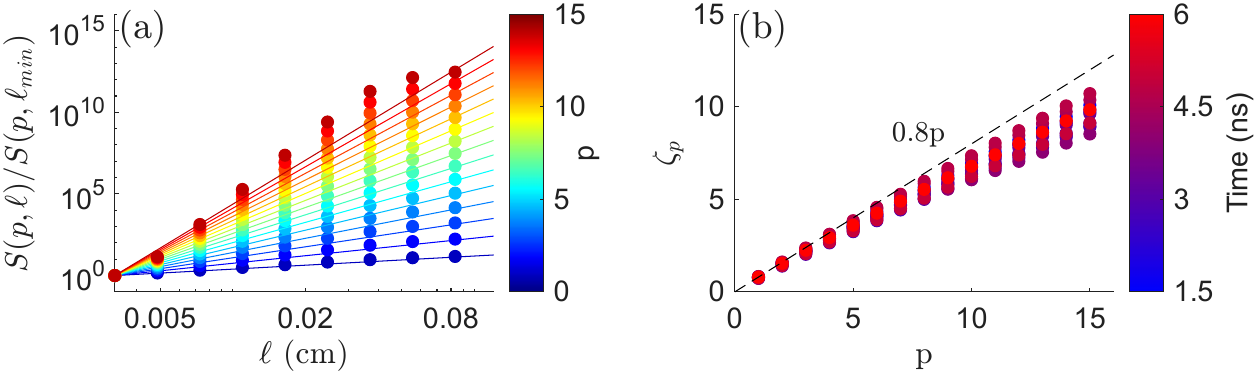} 
\caption{\label{structure_function} \textbf{Structure function characterization of the magnetic turbulence.} (a) Two-point correlation structure functions \cite{Davidson2015} $S(p,\ell) = \langle |\delta B_\ell|^p\rangle $ (colored dots), where we average on the PDF at various values of $\ell$, for the magnetic field map shown in Fig.~\ref{TurbuExample}b. The structure functions are fitted to a power law (solid lines), $S(p,\ell)\propto \ell^{\zeta_p}$. For visibility, the structure functions plotted are normalized by their value at the smallest increment $\ell$. (b) Scaling exponents $\zeta_p$ of space-averaged structure functions of the variation of magnetic field.  
}
\end{figure*}

To evaluate the structure function of the turbulence, we consider two-point correlation statistics \cite{Davidson2015} expressed as: 
 $S(p,\ell) = \langle |\delta B_\ell|^p\rangle $, where we average on the PDF at various values of $\ell$
\cite{
Frisch1995}. Fig.~\ref{structure_function}a shows structure functions corresponding to the magnetic field map shown in Fig.~\ref{TurbuExample}b. We note that the structure functions are predicted  to scale as a power law, $S(p,\ell)\propto \ell^{\zeta_p}$. Such fits correspond indeed quite well to the recorded structure functions, as shown in Fig.~\ref{structure_function}a. Fig.~\ref{structure_function}b shows the scaling exponents of space-averaged structure functions
of the variation of magnetic field, $\zeta_p$, as a function of structure function order $p$. 
In Kolmogorov turbulence, fluctuations are self-similar, so magnetic field increments scale $ \ell$ as $\delta B_{\ell} \sim \ell^{1/3}$, implying $S_p(\ell)\sim \ell^{p/3}$ and therefore a linear relation $\zeta_{p}=p/3$.  Intermittency breaks this self-similarity because the turbulence becomes dominated by strong rare structures such as current sheets or sharp magnetic bends. Across such structures, magnetic field increments can remain unusually large even at small separations, rather than weakening according to the Kolmogorov scaling.
Low-order structure functions are dominated by the many ordinary fluctuations and therefore remain close to Kolmogorov scaling. High-order structure functions increasingly emphasize the rare strong events, because large increments contribute disproportionately to high powers. Since these strong jumps remain comparatively strong at small scales, the high-order structure functions decrease more slowly as the separation $\ell$ decreases than predicted by the Kolmogorov scaling $\ell^{p/3}$. 
As a result, the scaling exponents $\zeta_{p}$ increase more slowly with $p$, producing the characteristic flattening of the $\zeta_{p}(p)$ relation that signals intermittency, as observed here.
We note that the $\zeta_p$ are higher than those recorded e.g. in solar flare observations \cite{Abramenko2003}.
Here we should note that, since the magnetic field maps are path integrated and projected in a plane, under statistical homogeneity along the line of sight, this reduces to $S^{2D}_p(\ell)<S^{3D}_p(\ell)$, i.e. $\zeta_p^{2D}>\zeta_p^{3D}$, which is indeed what we observe. In other words, it shows that line-of-sight averaging suppresses high-order moments of the increments. 

\begin{figure}[hbtp]
\centering
\includegraphics[trim={0.5cm 0cm 1cm 0cm},clip, width=0.95\textwidth]{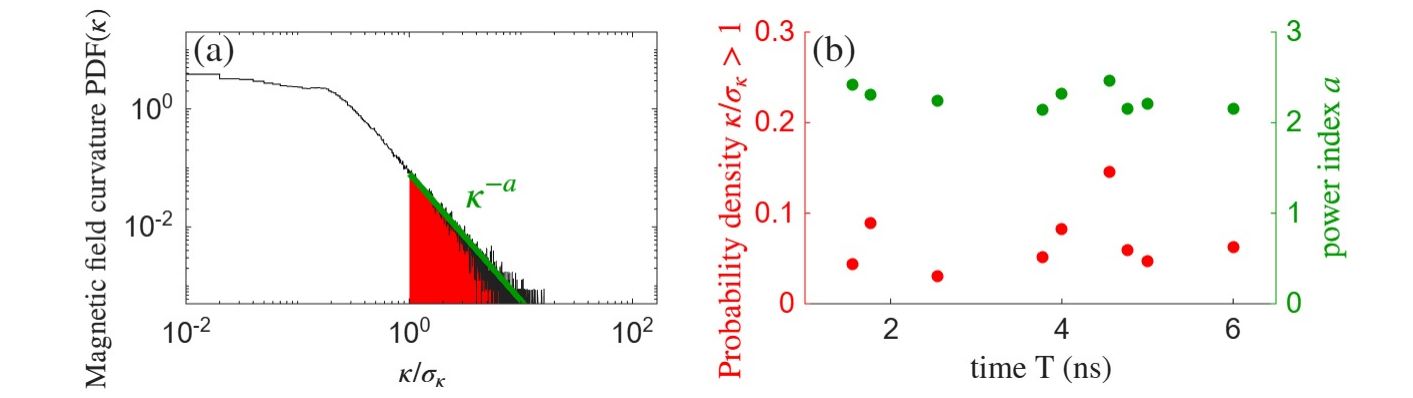} 
\caption{\label{CurvaturePDF} \textbf{Characterization of the magnetic field curvature over time.} (a) Probability function of the magnetic field curvature, for the magnetic field map shown in Fig.~\ref{TurbuExample}b. The area marked in red is the area for which $\kappa/\sigma_{\kappa}>1$. The green line is a power fit to the tail $\kappa/\sigma_{\kappa}>1$ of the curve. 
(b) Red - Probability of $\kappa/\sigma_{\kappa}>1$ as a function of time. It corresponds to the area delineated in red in (a). Green - power index of the curvature PDF tail in  the same zone.}
\end{figure}

To complement the analysis detailed above, we have also looked at another significant attribute of turbulence, which is the magnetic field curvature. To assess variations along the field, the dynamo literature indeed often considers 
 $\kappa=
 \frac{|(\vec{B}\cdot \vec{\nabla}) \vec{B}|}{B^2}$ \cite{Kempski2023}. In the definition of $\kappa$ presented here, the magnetic field vector $\vec{B}$ does not include the external guide field as the curvature is in the projected plane.
We constructed histograms of $\kappa$ normalized by its RMS value $\sigma_\kappa$. Such a histogram is shown in Fig.~\ref{CurvaturePDF}a for the field map presented in Fig.~\ref{TurbuExample}b.

For each such histogram computed from the magnetic field measured at various times, we calculate the area in which $\kappa/\sigma_{\kappa}>1$, which indicates the presence of structures with high curvature
. This area is indicated in red in Fig.~\ref{CurvaturePDF}a. This tail of the PDF $\kappa$, i.e. the portion where  $\kappa/\sigma_{\kappa}>1$, can be well fitted by a power fit ($PDF\sim\kappa^{-a}$), and that power fit result is shown in green in Fig.~\ref{CurvaturePDF}a. The dependence of the area as well as of the power index over time are both shown in Fig.\ref{CurvaturePDF}b, which again shows that the magnetic turbulence, apart from growing in strength (see Fig.\ref{Power Scaling}b) is  stable with respect to its statistical parameters  throughout the experiment. Interestingly, the power index, ${a}$, having value between 2 and 3, matches well recent simulations of large-amplitude turbulence \cite{Kempski2025
} and direct measurements of field-line curvature in the turbulent magnetosheath \cite{Bandyopadhyay2020}.

This demonstration that we can quantitatively produce astrophysically relevant (see Appendix \ref{secA2}) magnetic turbulence in the laboratory opens many doors for improving our understanding of many astrophysical phenomena. 
A significant interest of laboratory experiments, as demonstrated here, is that we can vary at play the level of turbulence. Further, we could also vary the Hall parameter, e.g. by changing the parameter of the turbulence-seeding laser, the seed magnetic field strength, and the density of the ambient medium (thus changing the ambient medium collisionality). Thus, this leads to the possibility of investigating the important question of shock-turbulence interactions, which is thought to play a key role in accelerating particles in the Universe. Indeed, we estimate that the Alfvénic Mach number for the turbulence  medium that can be laser-driven, as demonstrated here, can be in the range ~0.5-10, which is in good relation to estimated astrophysical values for the interstellar medium \cite{Vink2020} and for interplanetary shocks in the heliosphere \cite{Trotta2025}. Further, the flexibility of the present laboratory platform would  allow tuning at will the ratio between the turbulence characteristic scale and that of a shock that would be driven by an auxiliary laser \cite{Yao2021a}, to be propagated into the magnetic turbulence that we have here characterized.   Knowing all the energetics of the experiment, and the absolute number and spectra of the accelerated particles \cite{Yao2021a}, such platform would allow to quantify the absolute efficiency of particle acceleration in controlled conditions, which would then serve as a benchmark for global simulations used to analyze astrophysical observations \cite{Orlando2021}.  

Furthermore, the properties of the magnetic turbulence we have produced  also open the door to characterizing the transport of high-energy non-thermal charged particles  within an
intermittent magnetized turbulent plasma. Understanding the properties of such transport is of paramount significance in astrophysical environments, such as the interstellar, circumgalactic, and intracluster medium, where transport of high-energy cosmic rays can strongly aﬀect the evolution of supernova explosions, galactic winds, and black hole interactions with the ambient medium \cite{Ruszkowski2023}. Cosmic ray transport is indeed currently poorly constrained both theoretically and observationally, and present theories face challenges in explaining the observationally inferred transport properties in our Galaxy \cite{Kempski2022, Hopkins2022}.



%

\section{Methods}\label{sec11}

\subsection{\label{Experimental setup} Experimental setup}

The experiment was performed at the LULI2000 laser facility, located at  École Polytechnique (France)
. 
In the experiment, as shown in Fig.~\ref{ExperimentalSetup}, the turbulence was seeded by sending a high-power  (1.053~$\mu$m wavelength, 5 ns duration, 90 J energy, 1 mm focal spot top-hat diameter, resulting in an intensity of $2.3 \times 10^{12}$~$\mathrm{W/cm}^2$), through a random phase plate  \cite{Garnier1999-tu}, into a low-density ($5\times 10^{17}$~$\mathrm{cm}^{-3}$) H$_2$ ambient gas
. The ambient plasma was immersed in a $10$ T magnetic field that was homogeneous and steady state over the time scale of the experiment \cite{Albertazzi2013}. The magnetic field 
was oriented perpendicular to the turbulence driving beam.
An opening at the bottom of the magnetic-field driving coil allows inserting 
a ceramic pipe, from which the pulsed ambient medium 
was generated. Additionally, the coil has adequate openings, such that several auxiliary laser beams can be injected into it in order to diagnose the plasmas (see details below). The speckles within the 1 mm overall diameter beam have a diameter of 20~$\mu m$ and their statistical distribution in vacuum is well known \cite{Garnier1999-tu}.

In the initially
homogeneous ambient gas, the speckles of the randomized laser create pockets of heated
plasma, of various dimensions and thermal pressure, as set by the statistical distribution of the speckles
within the randomized laser. By generating a local high 
thermal pressure that 
overcomes that of the magnetic field, the speckles induce a local 
 outward compression of the frozen-in magnetic field in their immediate vicinity.
As the local heated plasma expands,  the neighboring compressed shells of B-field interact, overall forming a randomized net of
magnetic shells and voids (or “cells”).

The perturbations induced upon the initially homogeneous magnetic field by the speckled laser beam are diagnosed using proton radiography \cite{Schaeffer2023}. For this, broadband energy (in the MeV range) protons  
were generated using an auxiliary ultra-short laser (the PICO2000 laser, having 1 ps duration and  50~J energy on target, and focused to a spot of $\sim6$ $\mu m$ FWHM, leading to an on-target intensity of $6\times 10^{19}$ W/cm$^2$). The probing protons are accelerated from a secondary 20~$\mu m$ Al target, through the target normal sheath-acceleration (TNSA) mechanism \cite{Schaeffer2023}.
The proton beam is accelerated along the same axis as that of the external magnetic field. Following their propagation through the ambient plasma and the randomized magnetic field, they are collected 
onto a stack of radio-chromic films (RCFs) \cite{Schaeffer2023}.


The different layers in our RCF stack configuration that were used for magnetic field reconstruction allow to detect protons having energies of $E= 1.4,3.4$ and $4.7$~MeV. The distance between the proton source and the plasma was $5.05$~cm, and the distance between the plasma and the RCF stack was $4.75$~cm. The resulting magnification was thus $M=1.94$.

To characterize the plasma, additional measurements are performed, namely of its (i) density and (ii) temperature. For the density measurement, we used  optical interferometry, for which the plasma was probed by an auxiliary  laser, having a few mJ of energy, a 10-ns duration and a wavelength of 532 nm. The plasma was imaged onto Gated Optical Imagers (GOI) with opening gate times of 80 ps, allowing to grab snapshots of the plasma. The GOIs were timed at different moments, with respect to the turbulence-driving laser, to resolve the evolution of the plasma. For the temperature measurement, we used another auxiliary probe laser (3 ns duration, 15 J of energy, 526.5 nm wavelength). That latter laser was used to induce Thomson scattering (TS) onto the ion waves in the plasma \cite{Froula2012}. The scattered light was collected at 90° from the incident laser axis, such that the probing is performed in the collective regime, and was analysed using a high-resolution optical spectrometer coupled to a fast (15 ps resolution) streak camera. Details of the measurements are given below in Appendix \ref{density_temp}.

\subsection{\label{sec Proton radiography} Reconstruction of the magnetic field profiles using proton radiography}

Using the PROBLEM (PROton-imaged B-field nonlinear Extraction Module) algorithm \cite{Bott2017}, the path-integrated magnetic field is reconstructed. In the algorithm, the magnetic fields are retrieved by solving the logarithmic parabolic Monge-Ampère equation for the steady-state solution of the deflection field potential. The algorithm employs an adaptive mesh and a standard centered second-order finite difference scheme for spatial discretization, along with a forward Euler scheme for temporal discretization\cite {Sulman2011}. To infer the incident proton beam distribution, we apply a Gaussian blur filter to each image \cite{Bott2017}. Doing this removes all the fine gradients induced by the probed fields. We have selected the filter size to match, as close as possible, the proton flux distributions obtained in the absence of probed plasma, an example of which is shown in Appendix \ref{background_p}.

The use of the PROBLEM code to infer the magnetic field from the raw proton radiographic images is justified when the contrast regime is in the linear regime \cite{Schaeffer2023}, which is evaluated using the $\mu$-parameter \cite{Bott2017}. The $\mu$-parameter is the ratio of the deflections undergone by the protons to the characteristic length of the plasma. It is defined as: $\mu = \frac{\xi}{M l_B}$, where $\xi$ is the proton lateral displacement in the detector plane, $M$ is the magnification, and $l_B$ is the stochastic length of the path-integrated magnetic field. If $\mu > 1$, the proton trajectories would self-intersect. In this case, the system is non-injective, and several magnetic field configurations can produce the same proton radiography maps, preventing the utilization of the field reconstruction. In this case, the field that is reconstructed is the one with the minimal field value. In the data analyzed in this work, we estimate that  $\mu \le 1.58$ for the RCF layer that corresponds to a proton energy of $E=1.4$~MeV.
Examining the raw proton radiography images further shows that there are only a few structures that exhibit caustic features, which are characterized by splitting at the edge of magnetic structures. Thus, most of the image is in the linear regime, and the field can be reconstructed. Additionally, proton deflectrometry images analyzed in the caustic regime tend to have a spectral power law drop to the power of $P=2$ \cite{Bott2017}, which is different from our results. We should note that due to localized caustics in the images recorded at low proton energy, the proton radiography  underestimates the amplitude of the magnetic field. To address this, we corrected the magnetic field amplitude obtained form RCF layers probed with lower energies, by using the ratio of the probing proton velocities corresponding to the different RCF layers, thereby making the data from different layers consistent.

Since there is some unavoidable temporal jitter between the proton probing beam and the turbulence seeding laser, statistical quantities such as spectral index, moments of the PDFs, and magnetic field amplitude were averaged between data with similar sampling time, and the variations in these quantities is reflected  by the error bars in their respective figures.




\backmatter





\bmhead{Acknowledgments}
We thank the teams of the LULI2000 facility for their support. This research was supported in part by grant NSF PHY-2309135 to the Kavli Institute for Theoretical Physics (KITP).

\section*{Conflict of interest/Competing interests}

The authors declare that they have no conflict of interest regarding the research presented in the paper.

\section*{Data availability} 

All data needed to evaluate the conclusions in the paper are present
in the paper. Experimental data are archived on servers at the LULI laboratory and are available from the corresponding author upon reasonable request.

\section*{Code availability} 
 The PROBLEM code used to generate
Fig.~\ref{TurbuExample}b is available at https://github.com/flash-center/PROBLEM. The NEUTRINO code used to analyse the interferometry data is available at https://github.com/NeutrinoToolkit/Neutrino.

\section*{Author Contributions Statement} 
J.F. conceived the project. W.Y., N.M., J.B., P.-G.B., G.G., A.-M.T., Y.H., C.L. and J.F.performed
the experiments, with support from P.A., O.T. and
V.M. The data analysis was performed by I.C., W.Y., A.-F.-A.B., N.M., S.-N.C. and J.F., with discussions with D.C., E.d.H., A.M., M.R., P.K and O.A.  The bulk of the paper was written by I.C., 
S.N.C., O.A., M.R., P.K. and J.F. All authors commented and
revised the paper.









\begin{appendices}

\section{\label{density_temp} Background plasma density and temperature measurements}

\begin{figure*}[hbtp]
\centering
\includegraphics[trim={0 0 0 0},clip, width=1\textwidth]{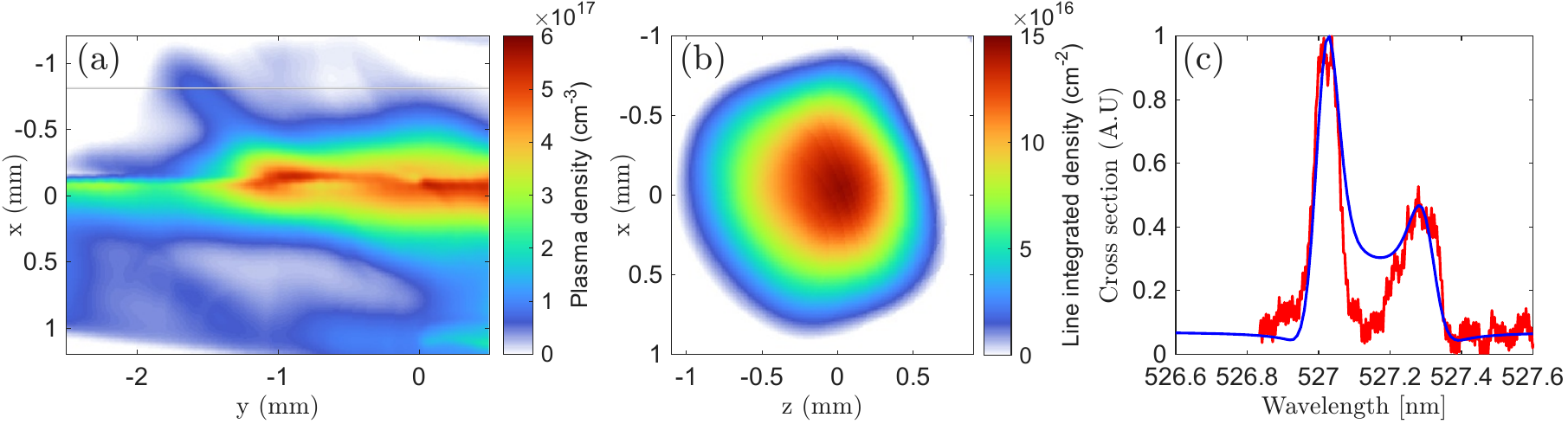}
\caption{\label{TS_inter} \textbf{Density and temperature characterization of the plasma.} a) Plasma density measured, using optical interferometry, in the xy-plane and 9~ns after the peak of the turbulence seeding beam. The line-integrated measured density  is Abel-inverted \cite{Vest1975} to retrieve the volumetric density. Since the plasma is not perfectly symmetric around the y-axis, Abel inversion is performed independently on the x-positive and x-negative half domains, and the results are shown in their respective domains, allowing to evaluate the asymmetry of the plasma.  b) Line integrated density measured, still using optical interferometry, 7~ns after the peak of the turbulence seeding beam and in the complementary xz-plane. c) Thomson scattering (TS) data measured 4.5~ns after the peak of the turbulence seeding beam. Raw TS measurement is in red (the signal falls to zero between the two ion peaks because of a black Al strip put on purpose to block the unshifted light) and fitted data in blue with $n_e=5\cdot10^{17} cm^{-3}$, $T_e=48.3 eV$, and $T_i=4 eV$. }
\end{figure*}

The results of the plasma density measurements, derived from optical interferometry, and of plasma temperature measurements, derived from 
TS, 
are presented in Fig.\ref{TS_inter}a-b) and c), respectively. 
The interferometry was performed along two different axes (i.e. 
along and perpendicular to the speckled beam laser), as shown, and analyzed using the Neutrino code \cite{Neutrino}.
Since the plasma measured 
in the x-z plane has an axis of symmetry, 
we were able to perform an Abel inversion \cite{Vest1975} of the raw line-integrated density map. 
Using the measurement 
in the x-z plane, which shows that the line integrated-density peaks at $\sim1.5\cdot10^{17}~ cm^{-2}$, and  the one in the x-y plane, which shows that 
the peak plasma density is $n_e \sim5\cdot10^{17} cm^{-3}$, we can infer  an estimate of the plasma length, along the y-axis, of $l_y=3~mm$
.
The fit of the TS scattering data fit yields plasma parameters of $n_e=5\cdot10^{17}~cm^{-3}$ $T_e=48.3~eV$ and $T_i=4~eV$. The Gaussian standard deviation of instrument was $0.282~nm$. The TS is in agreement with an inverse-Bremsstrahlung calculation of the speckled laser beam absorption in the plasma \cite{Yao2023}, which yielded a temperature of $T_e=46~eV$ for a density of $n_e=5\cdot10^{17}~cm^{-3}$.

Using these measured parameters, we estimate that the ratio of the electron Larmor radius to the electron collisional mean-free path to the  $2.2\cdot10^{-2}$, and that the ion Larmor radius to the ion collisional mean-free path is 1.8, meaning that the system is effectively magnetized and that collisional effects do not dominate its dynamics, similarly as in space and astrophysical systems. We can also assess that the plasma $\beta$, namely the ratio of plasma pressure to magnetic pressure, is $\beta=0.1$, similarly as in the solar corona.

\section{\label{background_p} Background proton beam profile}

\begin{figure}[hbtp]
\centering

\includegraphics[trim={0cm 0cm 0cm 0cm},clip, width=0.45\textwidth]{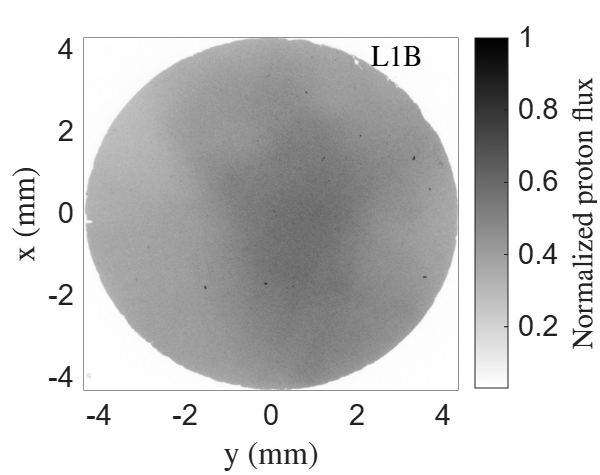} 
\caption{\label{No_B_Radiography} \textbf{Assessment of the background probing protons profile.} Proton radiography performed though the gas jet for protons of energy of $3.4$~MeV. The image shows the initial rather homogeneous flux distribution of the probing protons.}
\end{figure}

Fig.\ref{No_B_Radiography} shows the smooth profile of the probing proton beam as propagated from the source, here in the gas jet, in the absence of the external magnetic field and of the plasma.

\section{Full temporal characterization of the magnetic turbulence}\label{secA1}

\begin{figure*}[hbtp]
\centering
\includegraphics[trim={4cm 1cm 6cm 0cm},clip, width=1\textwidth]{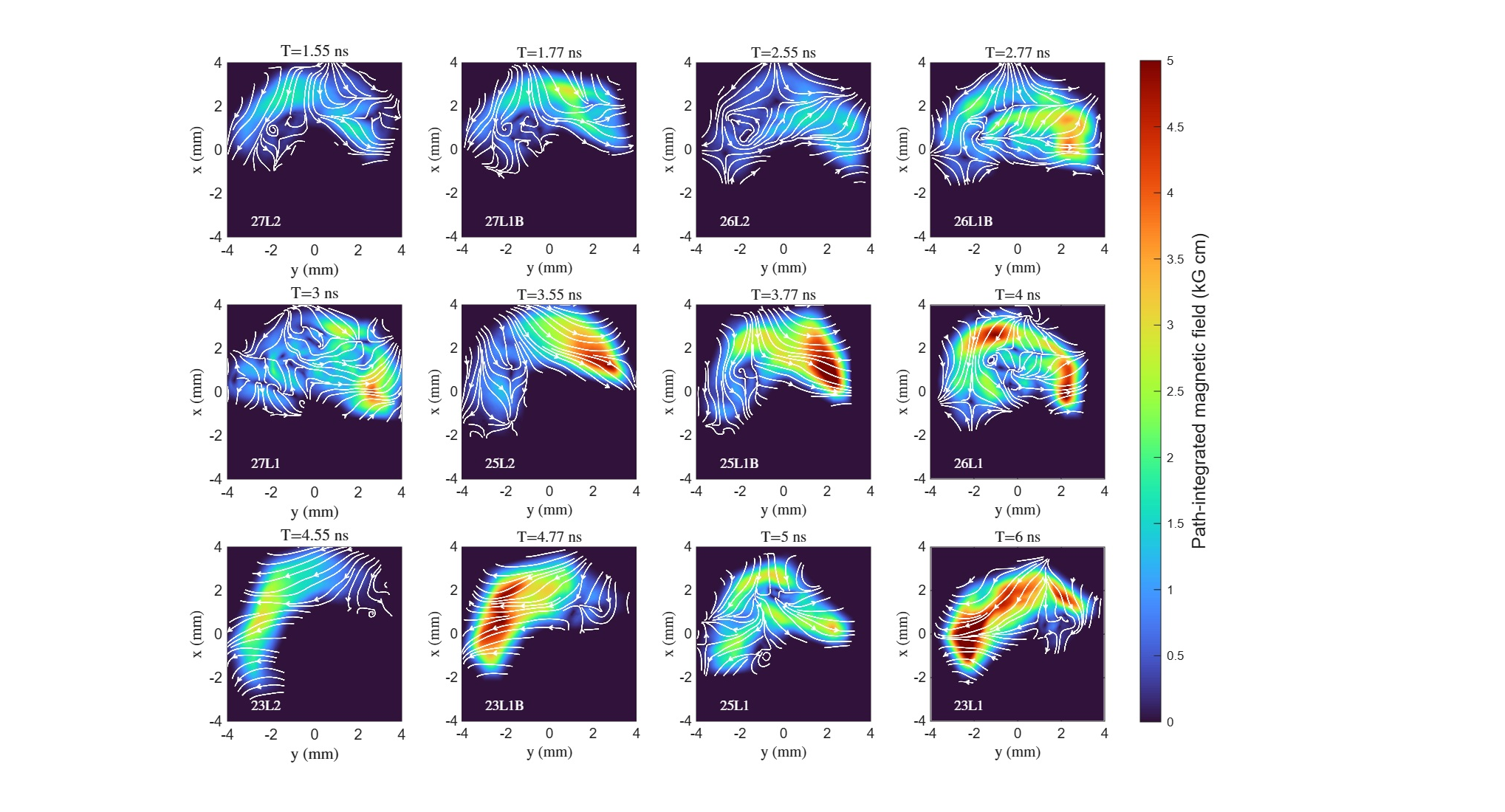}
\caption{\label{TurbuMaps} All the path-integrated magnetic field maps analyzed in this work from different laser shots. The corresponding probing times are stamped on each map along with the shot number and the RCF layer. The path-integrated magnetic field maps are retrieved using proton radiography using protons of energies of $E=~1.3,~3.4,~4.7$~MeV for RCF layers L1, L1B, L2 respectively, using the PROBLEM code \cite{Bott2017}; white arrows represent reconstructed magnetic field direction. }
\end{figure*}

Fig.~\ref{TurbuMaps} shows all the magnetic field maps analyzed from different shots and different RCF layers. We can note that 
the field maps (as the one shown in Fig.~\ref{TurbuExample}b) appear to be shaped as a "half-moon". This is due to the fact that the signal on the RCF stacks was contaminated by a Gaussian-like smear caused  by electrons and x-rays (which is roughly in the middle of the RCF films). This additional signal is observed to be constant on all the RCF layers of the same shot, which is evidence that it is not due to protons, thanks to the Bragg peak proton stopping power property. To avoid it interfering with the PROBLEM code analysis, we use the last layer in the stack of the RCF stack on which the proton signal is very weak to extract this noise, and  we subtract that signal from all the other layers. Additionally, we limited the region of interest of the analysis to include only the region where the protons are deflected, and to a region that avoids this noise area.  

\section{Scaling to astrophysical plasmas}\label{secA2}

In order to verify scalability of the plasma and magnetic field generated in the experiment with respect to astrophysical ones, we performed a scaling analysis.  Such a  dimensional analysis is needed for the identification of mutually independent dimensionless scaling parameters, and can be performed in different ways. 
Here are follow the analysis of  Ryutov \cite{Ryutov2018,Ryutov2000,Ryutov1999}, from which we can extract the scaling factors in space, density, and velocity between the laboratory plasma and a typical solar flare. 
In the laboratory we have a magnetic field of $B_{guide-lab}=10^5$~G, an ambient plasma density of $n_{lab}=5\cdot10^{17}$~cm$^{-3}$, and an electron temperature of $T_e=46$~eV, which corresponds to a thermal velocity of $V_{lab}=2840$~km/s.

For the solar flare, we take a temperature of $10^6$~K,  a flare length of $l_{astro}=10^6$~km,  a velocity of $V_{astro}=616$~km/s  
, and a typical density of $n_{astro}=10^{12}$~cm$^{-3}$ \cite{Aschwanden2008}. The scaling parameters between the two systems are thus as follows:

\begin{equation}
a=l_{astro}/l_{lab}=5\cdot10^{11} \\
\end{equation}
\begin{equation}
b=\rho_{astro}/\rho_{lab}=n_{astro}/n_{lab}=2\cdot10^{-6} \\
\end{equation}
\begin{equation}
c=V_{astro}/V_{lab}=0.21 \\ 
\end{equation}

We can now extract the scaling factor of the magnetic field strength between the laboratory plasma and the solar flare. 
$B_{guide-astro}/B_{guide-lab}=c\sqrt{b}=3\cdot10^{-4}$ corresponding to a field of $B_{guide-astro}=30$~G which is consistent with some solar flare measurements \cite{Karlicky2020,Kleint2010}.
The temporal scaling is given by  $a/c=2.3\cdot10^{12}$. Thus, a typical duration $t_{lab}\sim1$~ns is in correspondence to  $t_{astro}=38$~minutes, which is consistent with the observed time scale for the appearance of turbulence as induced by a solar flare \cite{Abramenko2003}.

\section{Evaluation of the predominance of magnetic-induced disturbances in the experimental protons radiographs
}\label{secA - BvsE}

\begin{figure}[htp]
    \centering
    \includegraphics[width=0.6\textwidth]{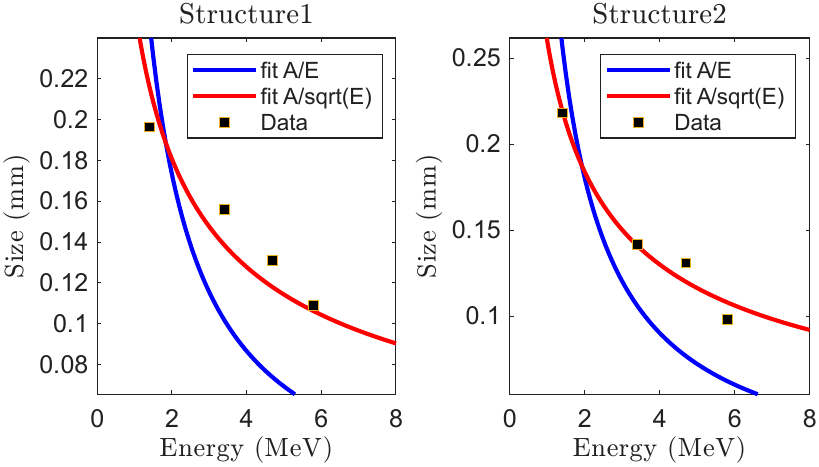}
    \caption{Scaling of proton radiography filaments from different RCF layers of the same shot, tracking the same structure on different films. The size of two filaments within the turbulence region, from the shot shown in Fig.\ref{TurbuExample}a, is analyzed at different proton probing energies. The data is fitted to deflection induced either by a magnetic force (in red) or an electric one (in blue). }
    \label{EnergyScaling}
\end{figure}

Both E- and B-fields can contribute to the proton dose modulations in proton radiographs. To show that the fields inducing proton deflections are of magnetic nature, we measured the size of a given filament structure within the plasma. This measurement was performed on different RCF layers of the same shot, identifying the same filament structure. We assume that the displacement of the protons is equal to the thickness of the filament. The different scaling of the displacement due to the magnetic Lorentz force ($1/\sqrt{E}$) and to the electric one ($1/E$) allows us to verify if the size of the filaments matches one of these scalings \cite{Bolanos2024}. These fits are shown in Fig.~\ref{EnergyScaling} for two given structures,  clearly showing that these deflections are due to magnetic forces. 

\end{appendices}


\bibliography{Mainbib}

\end{document}